# Gate-controlled weak antilocalization effect in inversion layer on p-type HgCdTe

Rui Yang, Guolin Yu, [*] Xinzhi Liu, Tie Lin, Shaoling Guo, Ning Dai, and Junhao Chu

*National Laboratory for Infrared Physics, Shanghai Institute of Technical Physics, Chinese Academy of Science, Shanghai 200083, People's Republic of China*

Yanfeng Wei, Jianrong Yang, and Li He

*Research Center for Advanced Materials and Devices, Shanghai Institute of Technical Physics, Chinese Academy of Sciences, 200083 Shanghai, China*

We discover weak antilocalization effect of two-dimensional electron gas with one electric subband occupied in the inversion layer on p-type HgCdTe crystal. By fitting the model of Iordanskii, Lyanda-Geller and Pikus to data at varies temperatures and gate voltages, we extract phase coherence and spin-orbit scattering times as functions of temperature and carrier density. We find that Elliot-Yafet mechanism and Nyquist mechanism are the dominating spin decoherence and dephasing mechanisms, respectively. We also find that the Rashba parameter is relatively large and the dependence of Rashba parameter upon carrier density is not monotonic and an optimal carrier density exists for the maximization of spin-orbit coupling.

HgCdTe has narrow energy gap and strong spin-orbit interaction.[1] It not only dominates the infrared techniques for decades[1] but also harbors many interesting spin-related properties[2] due to its strong spin-orbit coupling, thus has potentials in spintronics.[3] In the triangular potential well in the inversion layers of narrow-band semiconductors, Rashba spin-orbit coupling dominates the spin-splitting due to the lack of structure symmetry.[4, 5, 6] Rashba coupling includes contributions from electric field as well as boundary conditions, it provides possibility to manipulate spin-orbit coupling electrically and thus pave the way for spintronics.[4] Moreover, recently, some special narrow-band semiconductors and a HgCdTe-based quantum well have been recognized as topological insulator–a new kind of material in which strong spin-orbit interaction leading to formation of a topologically protected surface state.[7, 8, 9] This new material provides us promising vista of spintronics even topological quantum computation.[8, 10] Especially, among these materials, two-dimensional gas (2DEG) in HgTe/CdTe quantum well is the first experimentally-verified topological insulator,[9] and also is a 2-dimentional (2D) topological insulator which harbors quantum spin Hall effect.[7] Whether or not such topological insulator can be realized in the 2DEG confined in the triangular asymmetric potential well in the inversion layers of $Hg_{1-x}Cd_xTe$ (MCT) is a problem worth investigating. The investigation of spin-orbit coupling of $Hg_{1-x}Cd_xTe$ family material can advance our knowledge about Hg-based narrow-band semiconductors and benefits the field of spintronics as well as the emerging new field—topological insulator.

We report on experimental study of magnetoconductance of a two-dimensional electron gas formed in the inversion layer on p-type HgCdTe crystal. At high magnetic fields, we observe Shubnikov-de Haas (SdH) oscillations and integer quantum Hall effect (QHE), these are characteristic behaviors of an ideal two-dimensional electron gas with one subband occupied. From the thermal decaying of the amplitude of SdH oscillations, we get the effective mass of the ground electric subband. At low magnetic fields, prominent weak antilocalization effect is observed. After fitting the data at varies temperatures and carrier densities with the model developed by Iordanskii, Lyanda-Geller and Pikus (ILP),[11, 12] we investigate the phase coherence and spin-orbit scattering times as functions of temperature and carrier density. We find that the Rashaba parameter is quite large when compared with that of other materials and Nyquist and Elliot-Yafet (E-Y) mechanisms dominate dephasing and spin decoherence, respectively. We also find that the dependence of Rashba parameter upon carrier density is not monotonic and an optimal carrier density exists for the maximization of spin-orbit coupling.

The $5 \times 5$ mm$^2$ large p-type Hg$_{1-x}$Cd$_x$Te film with x= 0.218 used for the fabrication of our sample is grown on CdZnTe substrate by liquid-phase epitaxial (LPE) method. Anodic oxidation is used to form an inversion layer on the MCT film and a gate is made upon the inversion layer after deposition of rosin a few hundred μm thick as insulation layer (see inset of Fig. 1(b)). The sample is designated as 0225c here after. Indium is soldered onto our sample to facilitate Ohmic contacts. The magnetoresistance is measured in Van der Pauw configuration and the magnetic field is applied perpendicular to the film. A package of Keithley sourcemeters is used to measure magnetoresistance. All measurements are carried out in an Oxford Instruments $^4$He cryogenic system with temperature ranges from 1.3 to 9.0 K. A high-mobility 2DEG which dominates transport in the inversion layer as indicated by the SdH oscillation, quantum Hall effect (see Fig. 1(a), (b)) observed in experiments. At 1.4 K, the carrier concentration is $2.96 \times 10^{15}$ m$^{-2}$ and the mobility is 2.32 m$^2$V$^{-1}$s$^{-1}$.

In Fig. 1 (a) and (b), we can see excellent SdH oscillations and integer quantum Hall effect which are characteristic features of a high-quality 2DEG which dominates the transport at varies gate voltages. The shift of SdH oscillations and change of Hall curves are caused by the tuning of carrier density due to the gate. This reminds us of topological insulator realized in HgTe/CdTe quantum well, in which by tuning the carrier density with a gate, the 2DEG can evolve into a nominally insulator state which has a remaining conductance which is quantized in $e^2/2\pi\hbar$.[9]

The effective mass of ground electric subband is extracted by fitting the temperature dependence of the amplitude $A(T)$ of the SdH oscillation with the model[13] which describes the thermal decaying of SdH oscillations (see Fig. 1(c)),

$$\ln(\frac{A(T)}{T}) = c - \frac{2\pi^2 k_B m^*}{\hbar e B} T \qquad (1)$$

, where $c$ is a constant which doesn't depend on temperature, $e$ is the electronic charge, $\hbar$ is reduced Plank's constant, $k_B$ is the Boltzmann's constant, $m^*$ is the effective mass, $T$ is temperature, $B$ is magnetic field. The effective mass $m^*$ we get is $0.0185m_0$ ($m_0$ is the electron's mass in vacuum), this value is close to the effective mass of electrons at lowest subband in MCT inversion layers reported in other literatures.[14]

Clear weak localization effect with strong spin-orbit interaction has been observed in sample 0225c. In Fig.2, we can see that a negative magnetoconductance is imposed on a positive magnetoconductance background. The positive magnetoconductance background is caused by weak localization effect. The negative magnetoconductance 'peak' before the emerging of positive magnetoconductance is a mark of strong spin-orbit interaction and is known as weak antilocalization effect.[15] With the increase of temperature, we can see that the negative magnetoconductance 'peak' fades away gradually.

After the subtraction of conductivity background, the quantum correction of the magnetoconductivity at zero gate-voltage is fit according to ILP model:[11]

$$\Delta\sigma = \Delta\sigma(B) - \Delta\sigma(0) = \frac{G_0}{2}\left(F_t(b_\varphi, b_s) - F_s(b_\varphi)\right) \quad (2)$$

, where $G_0 = e^2/2\pi^2\hbar$; $b_\varphi = B_\varphi/B$; $b_s = B_s/B$; $B_\varphi = \hbar/4eD\tau_\varphi$; $B_s = \hbar/4eD\tau_{so}$; and $D$ is the diffusion constant and it's determined from Hall measurement and the resistivity at $B=0$. The meaning of $F_t(b_\varphi, b_s)$ and $F_t(b_\varphi, b_s)$ can be found in Ref. 11.

The fitting results are shown in Fig. 2 (The fitting is performed between $-B_{tr}$ to $B_{tr}$ with $B_{tr} \approx 8$ mT because ILP model is valid for $|B| < B_{tr} = \hbar/4eD\tau_{tr}$,[16] $\tau_{tr}$ is the transport scattering time). $\tau_\varphi$ ranges from $7.15 \times 10^{-12}$ to $2.06 \times 10^{-11}$ s, $\tau_{so}$ is approximately $4.52 \times 10^{-13}$ s. The Rashba parameter got from $\tau_{so}$, $1.02 \times 10^{-11}$ eVm, is close to values got in previous researches,[6, 17, 18, 19, 20, 21] this Rashba parameter is quite large when compared with that of other systems.[16, 22, 23] The large Rashba parameter is rooted in the relatively large ratio $\Delta_{so}/E_g$ (meaning of $\Delta_{so}$ and $E_g$ see Eq. (6)) of MCT when compared with other materials.

$\tau_\varphi$ decreases as temperature increases, see Fig. 3. From the relation between $\tau_\varphi$

and *T*, we can get the information about the dephasing mechanism. Generally speaking, $\tau_\varphi^{-1} \sim T^p$, *p*=1 and *p*=2, 3 are corresponding to Nyquist dephasing mechanism[24] and electron-phonon interaction dephasing mechanism respectively.[25, 26] For our data, the best fit is obtained for *p*=0.984 (see Fig.3, red line), thus the main dephasing mechanism is Nyquist mechanism.

The simplest model about the Nyquist dephasing mechanism[24] of Fermi liquid predicts a linear relation between $\tau_\varphi^{-1}$ and *T*:

$$\tau_\varphi^{-1} = \frac{1}{\alpha} \cdot T, \alpha = \frac{\frac{2\pi\hbar^2}{e^2}\sigma_0}{k_B \ln\left(\frac{\pi\hbar}{e^2}\sigma_0\right)} \quad (3)$$

$\alpha$ is the dephasing time, $\sigma_0$ is zero-field conductivity.

According to parameters got from linear fitting, we have dephasing rate $\alpha^{-1}$ ($3.43 \times 10^{10}$ s$^{-1}$), larger than theory value ($1.22 \times 10^{10}$ s$^{-1}$). The discrepancy indicates that Eq. (3) doesn't work very well here. In fact, this model only takes consideration the singlet contribution,[27] a more complete model which considers both singlet and triplet contributions has been developed,[27, 28]

$$\tau_\varphi^{-1} = (1+\frac{3(F_0^\sigma)^2}{(1+F_0^\sigma)(2+F_0^\sigma)})\frac{k_B T}{g\hbar}\ln(g(1+F_0^\sigma)) + \frac{\pi}{4}\left(1+\frac{3(F_0^\sigma)^2}{(1+F_0^\sigma)^2}\right)\frac{(k_B T)^2}{\hbar E_F}\ln(\frac{E_F \tau}{\hbar}) \quad (4)$$

, where $F_0^\sigma$ is the interaction constant for the triplet channel, g=$2\pi\hbar\sigma_0/e^2$ and $E_F$ is the Fermi energy.

The fitting according to Eq. (4) can be seen in Fig. 3 (green dash line), the best fit is reached when $F_0^\sigma$ =1.26, we'll see that this value is quite close to the value gotten below in the analysis of data at varies gate voltages. In addition, $F_0^\sigma$ also appears in the conductivity correction caused by electron-electron interaction,[29] and calculated temperature dependence of conductivity correction caused by electron-electron interaction based on $F_0^\sigma$ got above also agrees with experiment. Thus the Nyquist mechanism for Fermi liquid when taking both singlet contribution and triplet contribution into consideration is a good description of dephasing mechanism in our samples.

$\tau_{so}$ is basically temperature-independent, (see Fig. 3), this temperature-independence is relevant to the spin decoherence mechanism in MCT. Elliot-Yafet (EY) mechanism is believed to prevail in narrow-gap semiconductors such as HgCdTe we used here.[30] Our sample is in degenerate regime because $E_F$ is three magnitudes larger than $k_BT$. In degenerate 2D system, the temperature dependences of $\tau_{so}$ corresponding to EY mechanism is described by the following formula:[31,32]

$$\frac{1}{\tau_{so}^{EY}} \approx \left(\frac{\Delta_{so}}{\Delta_{so}+E_g}\right)^2 \left(1-\frac{m^*}{m_0}\right)^2 \frac{E_c E_k}{E_g^2} \frac{1}{\tau_p(T)} \quad (5)$$

, where $\Delta_{so}$ is spin-orbit splitting energy of valence band, $E_g$ is energy gap, $E_c$ is quantum confinement energy, $\tau_p$ is momentum scattering time, for degenerate semiconductors, $E_k=E_F$;[33] Thus in our MCT samples, $\tau_{so}^{-1}$ is temperature-independent. According to the observed value of $\tau_{so}^{-1}$, we can infer that $E_c \approx 165$ meV (we set $\Delta_{so}=1$ eV[14,19]), this value is consistent with previous researches concerning $E_c$ (~100 meV) in inversion layers of MCT.[14] In contrast, $E_c$ inferred from analysis based on the assumption that D'yakonov-Perel (DP) mechanism dominates is approximately 20 meV, which is too small, even less than $E_F$ ($\approx$ 40 meV), thus DP mechanism is excluded as the major spin-decoherence mechanism.

After the application of a gate-voltage, in Fig. 4, we can see that the weak antilocalization effect can be tuned by the gate. It means that the spin-orbit coupling can be manipulated by a gate and thus a promising landscape concerning spintronic devices based on electrically manipulated spin-orbit coupling is possible. After extracting $\tau_\varphi$ and $\tau_{so}$ at different gate voltages, we can investigate their dependence upon carrier density and conductivity thus clarify varies physics in our sample.

From the dependence of $\tau_\varphi^{-1}$ upon $\sigma_0$, see Fig. 5, we can decide the better one for the description of dephasing between Eq. (3) and Eq. (4). Fitting according to Eq. (4) is the blue curve in Fig. 5, the resulting $F_0^\sigma$ =0.978 is quite close to the value (1.26) from the analysis of temperature dependence of $\tau_\varphi^{-1}$ (see Fig. 3). The red curve in Fig. 5 is a plot related to Eq. (3), we can see that Eq. (4) gives the correct dependence. Thus the validity of model 4 which takes onto consideration both singlet contribution and triplet contribution in dephasing is further strengthened.

The dependence of spin-orbit scattering time, Rashba parameter and zero-field spin-splitting energy upon carrier density can also be readily gotten. We can see that the dependence of $\tau_p/\tau_{so}$ upon carrier density is linear, see Fig. 6(a), this follows the prediction of EY mechanism.[31, 32] The dependence of zero-field spin-splitting energy $\Delta_0$ upon carrier density is monotonic, see (Fig. 6(b)). Different from the monotonic dependence of $\Delta_0$ upon carrier density, a significant feature in the carrier dependence of Rashba parameter is that it's not monotonic (see Fig. 6(c)), we can see a peak emerges in the Rashba parameter ~ carrier density plot. This is because the spin-orbit scattering time monotonically decreases and the diffusion coefficient monotonically increases as the carrier density increases (see Fig. 6(d)) thus their product has a minimum, however, the Rashba parameter is inversely proportional to the product of spin-orbit scattering time and the diffusion coefficient, so, the Rashba parameter doesn't depend on carrier density monotonically and has a maximum. This is different from the dependence of Rashba parameter upon carrier density observed in other systems[4, 16, 34] where the Rashba parameter either doesn't depend on carrier density or monotonically decreases as carrier density increases but agrees with theoretical calculations concerning Rashba spin-splitting of MCT quantum wells performed by W. Yang and K. Chang.[35]

In summary, weak antilocalization effect in the inversion layers of HgCdTe films is discovered. After fitting the data at varies temperatures and gate voltages with ILP model, phase coherence and spin-orbit scattering times are extracted as functions of temperature and carrier density. We find that Nyquist dephasing mechanism is the dominating dephasing mechanism and EY mechanism is the dominating spin-decoherence mechanism. We also find that the dependence upon carrier density of Rashba parameter is nonmonotonic and an optimal carrier density exists for the maximization of spin-orbit coupling.

**Acknowledgements**

Authors of this work thank Longyuan Zhu, Guangyin Qiu for their help in the processing of samples. This work was supported by the Fund of Chinese Ministry of Personnel, the Special Funds for Major State Basic Research under Project No. 2007CB924901, and the Science and Technology Commission of Shanghai under Grant No. 07JC14059.

**Figure captions**

FIG. 1. (a) Magnetoresistance of 0225c at different gate voltages, we can see SdH oscillations. (b) Hall resistance at varies gate voltages, we can see quantum Hall plateaus clearly. Inset shows the configuration of our sample, region A (white) is p-type MCT film, region B (blue) is the inversion layer, C (black) is electrode, region D (yellow) is the insulation layer, region E (brown) is the gate. (c) ln($A(T)/T$) as a function of temperature. Blue squares are data, red line is fitting according to Eq. (1).

FIG. 2. Low-field magnetoconductance of 0225c, blue circles are the experimental data and red curves are fitting according to ILP model. The curves and circles have been shifted vertically for clarity.

FIG. 3. $\tau_\varphi^{-1}$ (blue circles) and $\tau_{so}^{-1}$ (green squares) as functions of temperature. Red curve is fitting result according to $\tau_\varphi^{-1} \sim T^p$, green dash line is fitting according to Eq. (4).

FIG. 4. Weak antilocalization effect at different gate voltages. Blue circles are the experimental data and red curves are fitting according to ILP model. The curves and circles have been shifted vertically for clarity.

FIG. 5. $\tau_\varphi^{-1}$ as a function of $\sigma_0$. Green blades are the experimental data and blue curve is fitting according to Eq. (4); red curve is plot related to Eq. (3).

FIG. 6. (a) $\tau_p/\tau_{so}$ as a function of $E_F$. Blue circles are the experimental data and red curve is linear fitting. (b) Zero-field spin-splitting energy $\Delta_0$ as a function of carrier density, blue triangles are experimental data. (c) Rashba parameter $\alpha$ as a function of carrier density. (d) $\tau_{so}$ and $D$ as functions of carrier density.